\documentclass[aps,prb,twocolumn,superscriptaddress,showpacs]{revtex4}
\usepackage{graphicx}
\usepackage{amsmath,amssymb}
\usepackage{times}
\usepackage{color}
\usepackage[normalem]{ulem}

\begin{document}

%Title of paper
\title{
Non-monotonic temperature dependence of thermopower in strongly correlated electron systems}
\author{M. Matsuo}\email[]{matsuo.mari@jaea.go.jp}
\affiliation{Advanced Science Research Center, Japan Atomic Energy Agency, Tokai 319-1195, Japan}
\affiliation{CREST, Japan Science and Technology Agency, Tokyo 102-0075, Japan}
\author{S. Okamoto}
\affiliation{Materials Science and Technology Division, Oak Ridge National Laboratory, Oak Ridge, Tennessee 37831, USA}
\author{W. Koshibae}
\affiliation{Cross-Correlated Materials Research Group (CMRG), RIKEN, Wako 351-0198, Japan }
\affiliation{CREST, Japan Science and Technology Agency, Tokyo 102-0075, Japan}
\author{M. Mori}
\affiliation{Advanced Science Research Center, Japan Atomic Energy Agency, Tokai 319-1195, Japan}
\affiliation{CREST, Japan Science and Technology Agency, Tokyo 102-0075, Japan}
\author{S. Maekawa}
\affiliation{Advanced Science Research Center, Japan Atomic Energy Agency, Tokai 319-1195, Japan}
\affiliation{CREST, Japan Science and Technology Agency, Tokyo 102-0075, Japan}

\begin{abstract}
We examine the temperature dependence of thermopower 
in the single band Hubbard model using dynamical-mean-field theory. 
The strong Coulomb interaction brings about the coherent-to-incoherent crossover 
as temperature increases.  As a result, 
the thermopower exhibits non-monotonic temperature dependence and 
asymptotically approaches values given by the Mott-Heikes formula.  
In the light of our theoretical result, 
we discuss the thermopower in some transition metal oxides.  
The magnetic field dependence of the thermopower is also discussed.  
\end{abstract}
\pacs{PACS numbers: 72.15.Jf, 71.10.Fd, 75.20.Hr}
\keywords{}
\maketitle
%%%%%%%%%%%%%%%%%%%%%%%%%%%%%%%%%%%%%%%%%%%%%%%%%

%%%%%%%%%%%%%%%%%%%%%%%%%%%%%%%%%%%%%%%%%%%%%%%%%
%\section{Introduction}

Thermopower is none other than 
the amount of entropy flow along with 
the electric current.\cite{maekawabook}  
The consideration on the entropy in thermodynamics 
tells us the low and high temperature $(T)$ limits of the thermopower:   
In the metallic systems, the thermopower goes to zero as $T\rightarrow 0$.  
On the other hand, the high-temperature limit of thermopower is given by 
the entropy consideration~\cite{maekawabook,chaikin,ktm} 
in the atomic limit.  
In the strongly correlated systems, 
the spin and orbital degrees of freedom 
enhance the high-temperature thermopower.\cite{maekawabook,chaikin,ktm}  

In the low-temperature limit, 
the ratio of the thermopower $Q$ and $T$ is proportional to 
the derivative of the density of states 
$D(\omega)$ with respect to the energy 
$\omega$ measured from the chemical potential $\mu$  
as $Q/T \propto-\partial D (\omega)/\partial\omega |_{\omega=0}$. 
By this relation, 
not only the sign but also the magnitude of the thermopower of 
conventional metals is well understood. 
The thermopower is a sensitive tool for the electronic states. 

The electron correlation brings about 
exotic electronic phases such as 
an anomalous metal 
near the Mott transition.  
In the vanadium oxide, La$_{1-x}$Sr$_x$VO$_3$, 
the filling control Mott transition is realized, 
and non-monotonic temperature dependence of the 
thermopower is observed.\cite{uchida}  
The temperature dependence manifests a crossover of 
coherent-to-incoherent charge transport.  
This phenomenon is common to transition metal oxides.  
The cobalt oxide, Na$_x$CoO$_2$ 
is an example.\cite{motohashi,foo,chou,ishida,wang,valla,hassan}   
In the photoemission spectroscopy measurements,\cite{valla,hassan} 
it is reported that 
the coherent motion of charge carriers 
is rapidly suppressed with increasing 
temperature, 
and the quasi-particle-peak disappears at $\sim$200 K.  

In this paper, 
we study the role of the strong Coulomb interaction on thermopower, 
whose temperature dependence is particularly examined in detail. 
For this purpose, 
the single band Hubbard model is adopted as a minimum model and 
the strong Coulomb interaction 
is treated in the dynamical mean field theory (DMFT)\cite{georges}, 
 which can capture the coherent-to-incoherent crossover due to the strong Coulomb interaction.  
This method based on the local picture is useful to understand the overall behavior of thermopower as a function of temperature.
We find that the Coulomb interaction significantly affects 
the temperature and magnetic field dependence of the thermopower.  
The Coulomb interaction is found to give rise to a 
non-monotonic temperature-dependence, 
which is well described by the entropy consideration at high temperatures.  
In the light of our theoretical results, 
we discuss the thermoelectric response 
in some transition metal oxides.

We start with the single band Hubbard model:
\begin{eqnarray}
H=\sum_{k \sigma} \varepsilon_k c_{k \sigma}^\dag c_{k \sigma}
+ U\sum_{i} n_{i\uparrow} n_{i\downarrow}
- \mu\sum_{i} (n_{i\uparrow}+n_{i\downarrow}),
\label{Hamiltonian}
\end{eqnarray}
where $\sigma(=\uparrow, \downarrow)$ denotes electron spin, 
$\varepsilon_k$ is the dispersion relation of the non-interacting electrons  
and other notations are standard.  
In the DMFT formalism, the resulting equations are the functions of the 
density of states for the non-interacting electrons.  
We denote the \lq\lq bare" density of states by $D_0$, 
and take the semicircular function of the energy $\varepsilon$, 
$D_0(\varepsilon)=[2/({\pi W^2})]\sqrt{W^2 - \varepsilon^2}$, 
with the normalization $W=1$ throughout this paper.  
Because the Hamiltonian has the particle-hole symmetry,  
all the results shown here are in the case that 
the electron concentration $n>1$.  
To solve the single impurity problem in DMFT, 
we employ the non-crossing approximation 
(NCA),\cite{maekawa,bickers} and the iterated perturbation theory 
(IPT),\cite{rozenberg,pruschke1,pruschke2,georges,kajueter,palsson,merino} which do not require the analytic continuation from the imaginary frequency axis.  
Using the calculated spectral density $\rho_\sigma (\varepsilon ,\omega )$ 
through DMFT, the thermopower $Q$ is expressed by, 
$Q=-(k_B/ e)(A_1/ A_0)$,
where
\begin{eqnarray}
A_l=\frac{\pi}{\hbar k_B}\int d\omega 
\frac{(\beta\omega)^l}
{4\cosh^2\left(\beta\omega /2\right)}
\sum_\sigma\int d\varepsilon\rho^2_\sigma (\varepsilon ,\omega )D_0(\varepsilon),
\label{a01}
\end{eqnarray}
where 
$\rho_\sigma (\varepsilon ,\omega )={\rm Im}[1/\{\omega+\mu-\varepsilon - \Sigma_\sigma (\omega)\}]$
with 
$\Sigma_\sigma (\omega)$ the electron self-energy.

 In the earlier studies, a saturation behavior of thermopower at high temperatures in the Hubbard model was discussed.\cite{om,shastryRev,phillips,palsson}   
We find in the following 
that the asymptotic behavior of $Q$ at 
high-temperatures 
shows a non-monotonic temperature dependence.  
%%%%%%%%%%%%%%%%%%%%%%%%%%%%%%%%%%%%%%%%%
\begin{figure}[b]
\begin{center}
\includegraphics[width=85mm,clip]{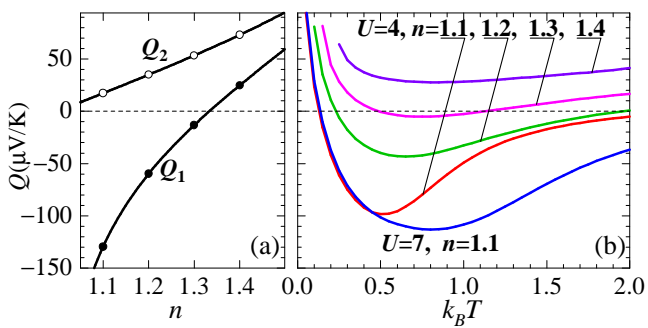}
\end{center}
\caption{(Color online)
(a) Thermopower in the high temperature limit, $Q_1$ and $Q_2$, vs. electron density, $n$. 
(b) Temperature dependence of the thermopower 
calculated by DMFT with the NCA impurity solver, 
for various parameter sets of $U$ and $n$.} 
\label{fig1}
\end{figure}
%%%%%%%%%%%%%%%%%%%%%%%%%%%%%%%%%%%%%%%%%
Before presenting the numerical results, 
let us note the high-temperature thermopower $Q$ 
of the model Eq.~(\ref{Hamiltonian}), on the entropy consideration.  
The independent variables of the function $Q$ are 
the electron concentration $n$, the Coulomb interaction $U$, and 
temperature $T$.  
For fixed $n$, we have {\it two} high-temperature limits:
 i) $Q_1:= Q(T\rightarrow\infty,U)$ by keeping $k_BT< U$.  
Because $U\rightarrow\infty$ is achieved before $T\rightarrow\infty$ in this case, 
$Q_1$ is given by,
\begin{eqnarray}
Q_1=\frac{k_B}{e}\ln\left(2\frac{n-1}{2-n}\right).
\label{q1}
\end{eqnarray}
ii) $Q_2:= Q(T\rightarrow\infty,U)$ by keeping $k_BT> U$.
In this limit, $U$ is of less importance, so that 
the result is written as the well known Heikes formula, 
\begin{eqnarray}
Q_2=\frac{k_B}{e}\ln\frac{n}{2-n}.
\label{q2}
\end{eqnarray}
Therefore, we expect the {\it two} different asymptotic behaviors, i.e., 
$Q_1$ and $Q_2$, 
and furthermore, a {\it sign-change} of the thermopower 
may occur in the temperature dependence:   
Figure \ref{fig1}(a) shows the $n$ dependence of 
the high-temperature limits of the thermopower, 
$Q_1$ and $Q_2$.  
For $n<1.3$, $Q_1$ is negative whereas $Q_2$ is positive.  

Figure \ref{fig1}(b) shows 
the temperature dependence of the thermopower 
calculated by DMFT with the NCA impurity solver.  
We find the non-monotonic temperature dependence of the thermopower.   
This is well understood as the asymptotic behavior with the 
high-temperature limits, $Q_1$ and $Q_2$:  
In the temperature region, 
$k_BT \agt 0.4$,
$Q$ is in the range between $Q_1$ and $Q_2$ for each $n$ 
(see the closed and open dots in Fig.~\ref{fig1}(a)).
With increasing 
$T$, 
$Q$ approaches $Q_1$ first, and further 
increasing 
$T$, $Q$ shows the saturation behavior 
given by $Q_2$.  
For $n=1.4$, $Q$ is always positive.  
But for $n=1.3$, $Q$ changes its sign twice with increasing temperature.  
With further decrease in $n$, the absolute value of the minimum of $Q$ 
is enhanced and becomes closer to $Q_1$.  
This is because the Coulomb interaction $U$ is more effective near 
half filling.  
The effect of the Coulomb interaction is made much clear by 
the $U$ dependence in $Q$.  
In Fig.~\ref{fig1}(b), the results of $U=4$ and 7 are shown for the same 
electron concentration, $n=1.1$.  
We see that the asymptotic approach of $Q$ to $Q_1$ is obvious for larger $U$.  
It is worth noting that  
the entropy consideration on the high-temperature thermopower works well even 
at finite temperatures.  

The single band Hubbard model Eq.~(\ref{Hamiltonian}) has 
2-fold degeneracy on the singly-occupied site.  
This is a disadvantageous condition for the NCA impurity solver 
because the approximation is based on an expansion in $1/N$ 
where $N$ is the ionic-angular-momentum degeneracy:  
For  
$k_BT \alt 0.4$, 
$Q$ increases rapidly with decreasing temperature.  
The numerical calculation simultaneously becomes unstable 
and eventually breaks down at certain temperature, 
i.e., the imaginary part of the calculated self-energy 
becomes positive at small frequencies.  
This will be improved for large $N$ systems, 
though further considerations are required for the thermodynamic properties 
near zero temperature.\cite{maekawa,bickers} 

%%%%%%%%%%%%%%%%%%%%%%%%%%%%%%%%%%%%%%%%%impurity solver
\begin{figure}
\begin{center}
\includegraphics[width=80mm,clip]{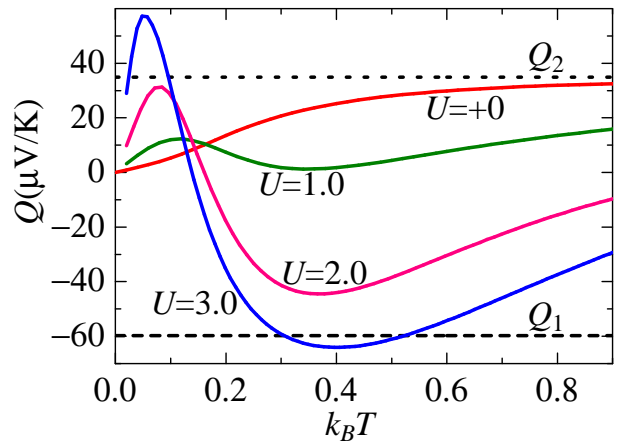}
\end{center}
\caption{(Color online)
Temperature dependence of the thermopower 
for $n=1.2$ calculated by DMFT with the IPT impurity solver.  
Broken and dotted lines are $Q_1$ and $Q_2$ for $n=1.2$, 
respectively.} 
\label{fig2}
\end{figure}
%%%%%%%%%%%%%%%%%%%%%%%%%%%%%%%%%%%%%%%%%

The thermopower is sensitive to the Coulomb interaction 
even for small $U$.  
To examine the thermopower in the small $U$ region, 
we use the IPT impurity solver for DMFT.  
In the calculation, we follow the method by Refs.~[\onlinecite{kajueter,merino}]
to obtain the self-energy.  
Figure \ref{fig2} shows the temperature dependence of $Q$ for $n=1.2$ 
and various $U$. In the small $U$ limit, $U=+0$, 
we use $D_0(\omega+\mu)$ for  
$\int d\varepsilon \rho_\sigma (\varepsilon ,\omega ) D_0(\varepsilon)$ in Eq.~(\ref{a01}).
In this case, $Q$ is a 
monotonically increasing function of $k_BT$:  
In the low temperature limit, 
$Q/T\propto-\partial D_0(\omega+\mu)/\partial\omega |_{\omega=0}$
and at high temperatures, 
$Q$ asymptotically approaches $Q_2$, i.e., the Heikes formula Eq.~(\ref{q2}).  
On the other hand, $Q$ shows the non-monotonic temperature dependence 
for finite $U$.  
Near zero temperature, the gradient of $Q$ with respect to $T$ 
is larger for larger $U$.  
With increasing 
$T$, $Q$ 
shows a maximum and then a minimum, and 
increases with further increasing 
$T$.  
For large $U$, 
the convergence of the numerical calculation in DMFT with the IPT impurity solver becomes poor,  
and the minimum of $Q$ grows beyond $Q_1$ (see the result for $U=3.0$ in Fig.~2).

We note the complementary characters of the impurity solvers NCA and IPT in DMFT.
The care must be taken to discuss the results of the thermopower shown in Figs.~\ref{fig1} and ~\ref{fig2}.
The NCA impurity solver is a perturbative expansion in powers of the hybridization between the impurity site and the effective bath.
Therefore, NCA is appropriate to discuss the thermopower for large $U$ 
(we find that the results are not plausible for $U < 3$).
On the other hand, the IPT impurity solver is a perturbative expansion with respect to the Coulomb interaction $U$
 and then appropriate for small $U$.
Further as an advantage, IPT can access the very low temperatures unlike NCA.
Next, we discuss the response of the low-temperature thermopower to the magnetic field by DMFT with the IPT impurity solver
focusing on the relatively small Coulomb interaction $U < 2W$ (total band width of the non-interacting system).

%%%%%%%%%%%%%%%%%%%%%%%%%%%%%%%%%%%%%%%%%
\begin{figure}
\begin{center}
\includegraphics[width=65mm,clip]{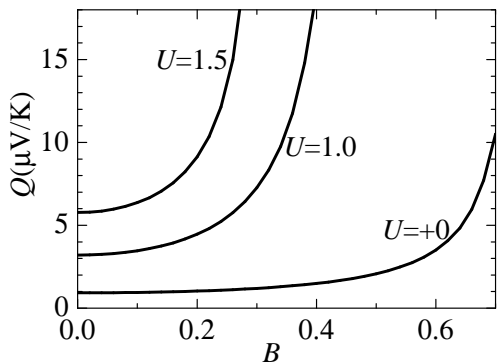}
\end{center}
\caption{Magnetic field dependence of the thermopower 
for $k_BT=0.02$ and $n=1.2$ calculated by DMFT with the IPT impurity solver.} 
\label{fig3}
\end{figure}
%%%%%%%%%%%%%%%%%%%%%%%%%%%%%%%%%%%%%%%%%

%%%%%%%%%%%%%%%%%%%%%%%%%%%%%%%%%%%%%%%%%
\begin{figure}[b]
\begin{center}
\includegraphics[width=65mm,clip]{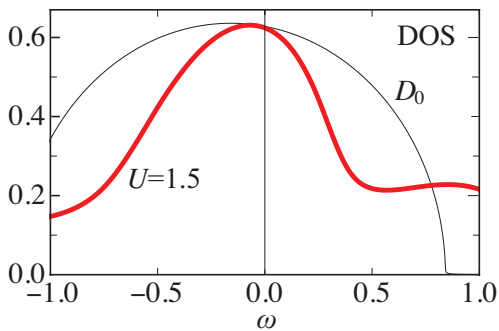}
\end{center}
\caption{
(Color online) The density of states (DOS) for $n=1.2$ at $k_BT=0.02$.
 Thick line is the result of DMFT with the IPT impurity solver for $U=1.5, B=0$. 
For reference, DOS for $U=0, B=0$ is presented by a thin line.
} 
\label{fig4}
\end{figure}
%%%%%%%%%%%%%%%%%%%%%%%%%%%%%%%%%%%%%%%%%

We introduce the Zeeman term, $-B\sum_i(n_{i\uparrow}-n_{i\downarrow})$,  
into the Hamiltonian Eq.~(\ref{Hamiltonian}) 
to discuss the response of the thermopower to the magnetic field $B$.  
Figure \ref{fig3} shows the magnetic field dependence of 
the thermopower for $k_BT=0.02$, $n=1.2$ and 
several values of Coulomb interaction $U$.  
We find that the Coulomb interaction enhances the thermopower and 
its response to magnetic field.  
The result for $U=+0$ is also presented as a reference 
to discuss the role of $U$ on the response of $Q$ to $B$.  
For finite $U$, the convergence of the numerical calculation becomes poor. 
Therefore, in Fig.~\ref{fig3}, we plot $Q$ within the range of $B$ where IPT is accessible.

In general, $Q$ in the low temperature region is well explained by the structure of density of states around the Fermi level.    
Hence, the increase of $Q$ by $B$ is also understood from the same viewpoint (see Fig.~\ref{fig4}).  
In the small magnetic field, 
we expand the interacting density of states $D_\sigma(\omega,B)$ with respect to $B$ as, 
\begin{eqnarray}
D_{\uparrow,\downarrow} (\omega,B)= D (\omega \pm B)
=\sum_{m=0}^\infty \frac{\partial^m D (\omega)}{\partial\omega^m}
(\pm B)^m. 
\label{dosbu}
\end{eqnarray}
Note that $D_\uparrow(\omega,B=0) = D_\downarrow(\omega,B=0) \equiv D(\omega)$. 
Since the total density of states, $D_{tot}(\omega)$, is written as 
$D(\omega+B)+D(\omega-B)$,
the derivative  
$\partial D_{tot}(\omega)/\partial\omega$ is expressed as 
\begin{eqnarray}
\frac{\partial D_{tot}(\omega )}{\partial\omega}=
2\sum_{m=0}^\infty \frac{\partial^{2m+1}D(\omega)}{\partial\omega^{2m+1}}
B^{2m}.
\label{dosb}
\end{eqnarray}
Through the relation 
$Q/T\propto-\partial D_{tot}(\omega)/\partial\omega |_{\omega=0}$
for the low-temperature thermopower,    
the magnetic-field dependence reflects the detailed structure of the density of states 
$D (\omega)$ near $\omega=0$.  
For $U=0$, the condition, semicircular density of state $D_0$ with $n=1.2$, results in 
an increase of $Q$ for small magnetic field, and 
the slow response to $B$ is a consequence of 
the non-interacting electron system.  
A simple dome structure of $D_0$ 
manifests the increase in $Q$ against $B$.    
Figure \ref{fig4} shows that the density of states for $U=1.5$ has a 
coherence peak near $\omega=0$.   
Therefore, as seen in Fig.~\ref{fig3}, the increase of $Q$ also appears.    

Equation (\ref{dosb}) suggests that 
the response of the low-temperature thermopower to magnetic field 
 reflects the detailed structure 
of the density of states near the Fermi level, 
i.e., the increase or decrease in the thermopower under the magnetic field is 
dependent on the differential coefficients, 
$\partial^{2m+1}D_{tot}(\omega)/\partial\omega^{2m+1}|
_{\omega=0}$. 
In this theoretical study, we use 
the single band Hubbard model  
with 
semicircular density of states $D_0$. 
The thermopower within DMFT is a function of the density of states.
Therefore, the increase in $Q$ by the magnetic field as shown in Fig.~\ref{fig3} is a consequence of the simple model 
 where  density of states has a negative slope at $\omega=0$ for $n=1.2$.
In reality, however, 
reflecting the band structure, the density of states near the Fermi level is 
certainly different from the simple dome shape.  
Here, the more important is that the response to the magnetic field 
is enhanced by the Coulomb interaction $U$.  
As shown in Fig.~\ref{fig4} by the thick line, the coherence peak is created by the Coulomb interaction accompanying the abrupt change in the density of states near the Fermi level.
As a result, the response of the thermopower to the magnetic field is enhanced as shown in Fig.~\ref{fig4}.

In the present study,  the qualitative behavior of the thermopower is clarified in the wide range of temperature, although we use the simplest model.    
An essential feature demonstrated here is that 
the strong Coulomb interaction brings about 
the large response of thermopower to external disturbance 
through the instability of the electronic state 
characterized by the narrow coherence peak.   
A measure of the instability is the renormalized energy scale, 
i.e., the width of the coherence peak.   
The renormalized or $reduced$ energy scale 
is of importance not only for the magnetic field dependence but also for 
the temperature dependence of the thermopower.  
In the temperature dependence, 
the entropy consideration for the high-temperature thermopower $Q_1$ 
works even  
at much lower temperatures than 
the band-width $2W$ and/or the Coulomb-interaction $U$
(see Figs.~\ref{fig1}(a) and \ref{fig2}).  
This means that the {\it high-temperature} 
is achieved on the basis of the $reduced$ energy scale 
by the Coulomb interaction.  

In the vanadium oxide, (La,Sr)VO$_3$, a non-monotonic temperature 
dependence of the thermopower is observed.~\cite{uchida} 
The thermopower approaches the two limiting values expected from the entropy consideration.  
This is the evidence of the coherent-to-incoherent crossover of 
the electronic states and well explained by our theory.  
In the cobalt oxide, Na$_x$CoO$_2$, 
a strongly renormalized quasi-particle band,  
which disappears near the room temperature, is observed by 
the photoemission spectroscopy measurements.~\cite{valla,hassan} 
As discussed in this paper, when electrons lose their coherence by increasing temperature, the thermopower simultaneously approaches to the asymptotic value obtained by the entropy consideration in the correlation dominant regime, i.e., $Q_1$ for $k_BT<U$. 
Hence, for the large thermopower of this material observed near the room temperature, the strong Coulomb interaction must be one of important factors. 
For more qualitative studies, one should employ the developed DMFT analysis such as cluster-DMFT\cite{MaierRMP} and
(cluster-)DMFT combined with the density functional theory.\cite{kotliarRMP}
In fact, the recent cellular DMFT studies with realistic band structures suggest that the non-local correlations improve the results.\cite{cellularDMFT} 
Yet, we believe that our study provides a good starting point for the future theoretical studies based on the advanced methods.  
On the other hand, in the partially-occupied $t_{2g}$ states, the spin-orbit coupling is not fully quenched 
as discussed in Refs.~\onlinecite{khaliullin} and \onlinecite{xiang}.  
In electronic systems with heavier elements such as Sr$_2$IrO$_4$,  
the spin-orbit coupling becomes comparable to the Coulomb interaction and the kinetic energy and, thus, 
plays a significant role 
for the Mottness of correlated systems.\cite{kim,jackeli,watanabe}  
This effect may be another path to a new thermo-electronics 
based on the strongly correlated electron systems.  

In summary, we have theoretically studied
the role of the Coulomb interaction in the thermopower.  
To clarify the role, we 
consider the single-band Hubbard model, and calculate the thermopower 
using the dynamical mean field theory.  
We find the non-monotonic temperature dependence of the thermopower. 
This is well described by the asymptotic form in the high-temperature limit given by the entropy consideration.     
The Coulomb interaction plays an important role to obtain such asymptotic behavior in a rather low-temperature region by creating the narrow quasi-particle band.

After completing the manuscript, 
we have noticed a recent work by W. Xu et al.\cite{xu} 
who studied the high-frequency limit of the thermopower in the strongly correlated system. 
They also reported the non-trivial temperature dependence. 

The authors are grateful to V. Zlati${\rm \acute{c}}$ for useful discussions.
This work is partly 
supported by Grants-in-Aid for Scientific Research from MEXT 
(Grant No. 19204035, No. 21360043, No. 22102501), 
the "K" computer project of Nanoscience Program, JST-CREST,   
and FIRST-Program. 
S.O. was supported by the U.S. Department of Energy, Office of Basic Energy Sciences, Materials Sciences and Engineering Division.


\begin{thebibliography}{99}
\bibitem{maekawabook}
As a review, 
S. Maekawa, T. Tohyama, S. E. Barnes, S. Ishihara, W. Koshibae, and G. Khaliullin, 
{\it Physics of Transition Metal Oxides}, vol. 144 of 
{\it Springer Series in Solid-State Sciences}, Springer (2004). 

\bibitem{chaikin}
P. M. Chaikin and G. Beni, Phys. Rev. B {\bf 13}, 647 (1976).

\bibitem{ktm}
W. Koshibae, K. Tsutsui, and S Maekawa, Phys. Rev. B {\bf 62}, 6869 (2000).

\bibitem{uchida}
M. Uchida, K. Oishi, M. Matsuo, W. Koshibae, Y. Onose, M. Mori, 
J. Fujioka, S. Miyasaka, S. Maekawa, and Y. Tokura, 
 Phys. Rev. B.  {\bf 83}, 165127 (2011).

\bibitem{motohashi}
T. Motohashi, R. Ueda, E. Naujalis, T. Tojo, I. Terasaki, 
T. Atake, M. Karppinen, and H. Yamauchi, Phys. Rev. B {\bf 67}, 064406 (2003). 

\bibitem{foo}
M. L. Foo, Y. Wang, S. Watauchi, H. W. Zandbergen, T. He, R. J. Cava, 
and N. P. Ong, Phys. Rev. Lett. {\bf 92}, 247001 (2004).

\bibitem{chou}
F. C. Chou, J. H. Cho, and Y. S. Lee, Phys. Rev. B {\bf 70}, 144526 (2004).

\bibitem{ishida}
Y. Ishida, H. Ohta, A. Fujimori, and H. Hosono, 
J. Phys. Soc. Jpn. {\bf 76}, 103709 (2007).

\bibitem{wang}
Y. Wang, N. S. Rogado, R. J. Cava and N. P. Ong, 
Nature (London) {\bf 423}, 425 (2003). 

\bibitem{valla}
T. Valla, P. D. Johnson, Z. Yusof, B. Wells, Q. Li, S. M. Loureiro,
R. J. Cava, M. Mikami, Y. Mori, M. Yoshimura, and T. Sasaki, 
Nature (London) {\bf 417}, 627 (2002).

\bibitem{hassan}
M. Z. Hasan, Y.-D. Chuang, D. Qian, Y. W. Li, Y. Kong,
A.P. Kuprin, A. V. Fedorov, R. Kimmerling, E. Rotenberg,
K. Rossnagel, Z. Hussain, H. Koh, N. S. Rogado, M. L. Foo, and
R. J. Cava, Phys. Rev. Lett. {\bf 92}, 246402 (2004).

\bibitem{georges}
A. Georges, G. Kotliar, W. Krauth, and M. J. Rozenberg, 
Rev. Mod. Phys. {\bf 68}, 13 (1996). 

\bibitem{maekawa}
S. Maekawa, S. Kashiba, M. Tachiki and S. Takahashi, 
J. Phys. Soc. Jpn. {\bf 55}, 3194 (1986).

\bibitem{bickers}
N. E. Bickers, Rev. Mod. Phys. {\bf 59}, 845 (1987).

\bibitem{rozenberg}
M. J. Rozenberg, X. Y. Zhang, and G. Kotliar, 
Phys. Rev. Lett. {\bf 69}, 1236 (1992). 

\bibitem{pruschke1}
Th. Pruschke, D. L. Cox, and M. Jarrell, Phys. Rev. B {\bf 47}, 3553 (1993).

\bibitem{pruschke2}
Th. Pruschke, M. Jarrell, and J. K. Freericks, Adv. Phys. {\bf 44}, 187 
(1995).

\bibitem{kajueter}
H. Kajueter and G. Kotliar, Phys. Rev. Lett. {\bf 77}, 131 (1996).

\bibitem{merino}
J. Merino and R. H. McKenzie, Phys. Rev. B {\bf 61}, 7996 (2000).

\bibitem{palsson}
G. P{\'a}lsson and G. Kotliar, Phys. Rev. Lett. {\bf 80}, 4775 (1998).

\bibitem{shastryRev}
B. S. Shastry, Rep. Prog. Phys. {\bf 72}, 016501 (2009).

\bibitem{phillips}
S. Chakraborty, D. Galanakis, and P. Phillips, Phys. Rev. B {\bf 82}, 214503 (2010).

\bibitem{om}
A. Oguri and S. Maekawa, Phys. Rev. B {\bf 41}, 6977 (1990). 

\bibitem{MaierRMP}
T. Maier, M. Jarrell, Th. Pruschke, and M. H. Hettler,
Rev. Mod. Phys. {\bf 77}, 1027 (2005).

\bibitem{kotliarRMP}
G. Kotliar, S. Y. Savrasov, K. Haule, V. S. Oudovenko, O. Parcollet,
and C. A. Marianetti,
Rev. Mod. Phys. {\bf 78}, 865 (2006).

\bibitem{cellularDMFT}
O. E. Peil, A. Georges, and F. Lechermann, arXiv:1107.4374.

\bibitem{khaliullin}
G. Khaliullin, W. Koshibae, and S. Maekawa, Phys. Rev. Lett. {\bf 93}, 176401 (2004).

\bibitem{xiang}
H. J. Xiang and D. J. Singh, Phys. Rev. B {\bf 76}, 195111 (2007).

\bibitem{kim}B. J. Kim, H. Jin, S. J. Moon, J.-Y. Kim, B.-G. Park, C. S. Leem, J. Yu, T.W. Noh, C. Kim, S.-J. Oh, J.-H. Park, V. Durairaj, G. Cao, and E. Rotenberg, 
Phys. Rev. Lett. {\bf 101}, 076402 (2008).

\bibitem{jackeli}
G. Jackeli and G. Khaliullin, Phys. Rev. Lett. {\bf 102}, 017205 (2009).

\bibitem{watanabe}
H. Watanabe, T. Shirakawa, and S. Yunoki, Phys. Rev. Lett. {\bf 105}, 216410 (2010).

\bibitem{xu}
W. Xu, C. Weber, and G. Kotliar, Phys. Rev. B {\bf 84}, 035114 (2011).

\end{thebibliography}
\end{document}